\documentclass[9pt,twocolumn,twoside]{osajnl}

\journal{jocn} 
\usepackage{acronym}
\usepackage{amssymb}
\usepackage{multirow}
\newcommand{\added}[1]{{\color{black}#1}}
\setboolean{shortarticle}{false}

\acrodef{EDFA}{Erbium-Doped Fiber Amplifier}
\acrodef{VOA}{Variable Optical Attenuator}
\acrodef{SELU}{Scaled Exponential Linear Unit}
\acrodef{NN}{Neural Network}
\acrodef{OSaaS}{Optical Spectrum as a Service}
\acrodef{ReLU}{Rectified Linear Unit}
\acrodef{MSE}{Mean Squared Error}
\acrodef{MAE}{Mean Absolute Error}
\acrodef{TL}{Transfer Learning}
\acrodef{ML}{Machine Learning}
\acrodef{SNN}{Self Normalizing Neural Networks}
\acrodef{WDM}{Wavelength Division Multiplexing}
\acrodef{DWDM}{Dense Wavelength Division Multiplexing}
\acrodef{WSS}{Wavelength Selective Switch}
\acrodef{ITU}{International Telecommunication Union}
\acrodef{ROADM}{Reconfigurable Optical ADD-DROP Multiplexer}
\acrodef{OCM}{Optical Channel Monitor}
\acrodef{PD}{Photo Diode}
\acrodef{PM}{Power Monitor}
\acrodef{SNR}{Signal-to-Noise Ratio}
\acrodef{OSNR}{Optical SNR}
\acrodef{GFF}{Gain Flattening Filter}
\acrodef{ILA}{In-Line Amplifiers}
\acrodef{EDF}{Erbium Doped Fiber}
\acrodef{AGC}{Automatic Gain Control}
\acrodef{QoT}{Quality of Service}
\acrodef{CDF}{Cumulative Distribution Function}
\acrodef{FLOP}{Floating Point Operation}
\acrodef{TFLOPS}{Tera Floating Point Operations per Second}
\acrodef{SHB}{Spectral Hole Burning}

\title{A Generalized Few-Shot Transfer Learning Architecture for Modeling EDFA Gain Spectrum (Invited FNWF 2024)}

\author[1]{Agastya Raj}
\author[2]{Zehao Wang}
\author[2]{Tingjun Chen}
\author[3]{Dan Kilper}
\author[1]{Marco Ruffini}

\affil[1]{CONNECT Centre, School of Computer Science and Statistics, Trinity College Dublin, Ireland}
\affil[2]{Duke University, Department of Electrical and Computer Engineering, Durham, NC, USA}
\affil[3]{CONNECT Centre, School of Engineering, Trinity College Dublin, Ireland}

\affil[*]{rajag@tcd.ie}

\begin{abstract}
Accurate modeling of the gain spectrum in Erbium-Doped Fiber Amplifiers (EDFAs) is essential for optimizing optical network performance, particularly as networks evolve toward multi-vendor solutions. In this work, we propose a generalized few-shot transfer learning architecture based on a Semi-Supervised Self-Normalizing Neural Network (SS-NN) that leverages internal EDFA features—such as VOA input/output power and attenuation—to improve gain spectrum prediction. Our SS-NN model employs a two-phase training strategy comprising unsupervised pre-training with noise-augmented measurements and supervised fine-tuning with a custom weighted MSE loss. Furthermore, we extend the framework with transfer learning (TL) techniques that enable both homogeneous (same-feature space) and heterogeneous (different-feature sets) model adaptation across booster, preamplifier, and ILA EDFAs. To address feature mismatches in heterogeneous TL, we incorporate a covariance matching loss to align second-order feature statistics between source and target domains. Extensive experiments conducted across 26 EDFAs in the COSMOS and Open Ireland testbeds demonstrate that the proposed approach significantly reduces the number of measurements requirements on the system while achieving lower mean absolute errors and improved error distributions compared to benchmark methods. 
\end{abstract}

\setboolean{displaycopyright}{false} 

\begin{document}

\maketitle
\footnote{This is a preprint of a paper accepted and published in the Journal of Optical Communications and Networking (JOCN). The final published version is available at: https://doi.org/10.1364/JOCN.560987}

\section{Introduction}
\label{section1}
The relentless growth of global data traffic, driven by emerging technologies such as 5G/6G, smart cities, autonomous systems, and the Internet of Things (IoT), has placed increasing demands on optical networks to sustain higher bandwidth, lower latency, and enhanced reliability \cite{fayad6GOpticalFronthaul2025}. These requirements are further increased by the industry-wide shift toward access-metro-core convergence, an architectural framework that integrates disaggregated network layers, optimizes spectral and energy efficiency through software-defined control, and enables seamless scalability across edge, metro, and long-haul domains \cite{ruffiniAccessMetroNetwork2017}. A critical component that enable these advancements is the \ac{EDFA}, which compensates for signal loss over long distances. This has an impactful role in optimizing optical network performance as the output power of the \ac{EDFA} determines launch power level of each optical channel, which further affects the magnitude of nonlinear impairments in fiber.  The noise figure of the \ac{EDFA}, which quantifies the degradation of \ac{SNR} due to amplification, has a direct impact on key performance metrics such as the \ac{OSNR} and \ac{QoT} \cite{mahajanModelingEDFAGain2020, yankovSNROptimizationMultiSpan2021}. The performance of \acp{EDFA} is inherently tied to wavelength-dependent gain profile, which exhibits complex nonlinear behavior under varying operating conditions, including pump power, channel loading, and operating mode \cite{jonesSpectralPowerProfile2023, liuModelingEDFAGain2021a}. Recently, increasing demands of high-bandwidth, low-latency applications has led to development of \ac{OSaaS}, which allows multiple parties to utilize the same network infrastructure, thereby increasing spectral utilization. However, in such a dynamic scenario with multiple stakeholders, predicting the system behavior is necessary to optimize performance and minimize security risks~\cite{osaas_ecoc_conf,Raj:25}. Therefore, accurate modeling of \ac{EDFA} gain spectrum is critical for optimizing physical-layer control in dynamic networks, particularly as operators reduce design margins to improve spectral efficiency.


\textcolor{black}{
Traditional \ac{EDFA} gain models rely on physics-based frameworks derived from the quantum-mechanical principles of \ac{EDF} dynamics, or on simplified expressions for \ac{AGC} mode, which requires only a limited amount of measurements \cite{agrawalFiberOpticCommunicationSystems2012}. However, these models are constrained by oversimplified assumptions (e.g., idealized two-level erbium systems, constant gain spectrum shape under all channel configuration), and neglect dynamic effects, manufacturing variability, and aging factors, reducing accuracy in practical deployments. }\added{Under dynamic channel loading conditions, the inaccuracy is further exacerbated for three reasons: 
\begin{enumerate}
    \item Critical \ac{EDFA} parameters~(e.g. $Er^{3+}$ concentration, fiber length) are rarely available for commercial units and are difficult to measure accurately~\cite{wangOpenEDFAGain2023}
    \item Commercial amplifiers exhibit complex multi-stage architectures that include gain-flattening filters, photo-detectors, couplers, and \acp{VOA} elements that are typically emitted in two-stage analytical models~\cite{lumentum_greybox}.
    \item Physical models struggle to reproduce \ac{SHB} behavior~\cite{de2023experimental}. 
\end{enumerate}

Recently, \ac{ML}-based \ac{EDFA} models have been proposed to overcome these limitations by collecting measurements from \ac{EDFA} devices~\cite{dt_book_chapter}. Several methods have been proposed for predicting the \ac{EDFA} gain spectrum under varying input power levels and channel loading configurations \cite{zhuMachineLearningBased2018,darosMachineLearningbasedEDFA2020}. Despite these advancements, comparing reported \ac{ML} models is difficult because the literature uses different experimental settings, measurement resolutions, and train-test splits. To mitigate the high measurement requirements of a purely data-driven approach, hybrid ML models have been proposed which combine physical \ac{EDFA} models with \ac{ML} techniques achieving a favourable accuracy versus measurement trade-off~\cite{zhuHybridMachineLearning2020}. Despite these innovations, even the best performing \ac{ML} techniques often require extensive labeled datasets that are impractical to collect on a live network and are typically trained on single-vendor devices, limiting their generalisability. 

These data and generalisation constraints become even more pronounced once individual amplifier models are stitched together to predict end-to-end system behavior. When per-component \ac{EDFA} models are cascaded to form a multi-span optical link model, uncompensated error accumulation can severely degrade prediction accuracy~\cite{wang2022optical, you2019osnr}. Two broad adaptive strategies have been investigated: (1) \textit{Parameter Refinement} methods that fine-tune physical model parameters after deployment but only partly correct the accumulated error~\cite{yankovSNROptimizationMultiSpan2021}; and (2) \textit{Cascaded Learning} methods, which treat the entire link as a end-to-end differentiable chain of component models and use sparse link-level measurements to back-propagate corrections, thereby reducing the link error to the level of a single component~\cite{chenSoftwareDefinedProgrammableTestbed2022,raj2024multi}. Cascaded learning effectiveness thus relies on a high-fidelity component model of the \ac{EDFA} gain profile, even though some margin for component-level error remains. 


Building and maintaining such high-fidelity models for every amplifier in a large, heterogeneous network is, however, both computationally and measurement-intensive. \ac{TL} is a promising path to reduce data dependency by leveraging knowledge from source domains to accelerate learning in target domains \cite{zhuangComprehensiveSurveyTransfer2021a}. Specifically, a well-characterized \ac{EDFA} model can be used to model other devices with only limited additional measurements. Recent research showed that \ac{TL} can be effective for modeling same-type \acp{EDFA} (e.g., booster-to-booster), achieving accurate predictions with just 0.5\% of the target dataset \cite{Wang:23}. However, the application of \ac{TL} across different amplifier types and vendors (e.g., booster and pre-amplifier) remains under-explored. Particularly in large networks comprising \acp{EDFA} from multiple vendors, \ac{TL} can offers a viable means to characterize the power dynamics using minimal additional data. In addition, existing models only consider the external features such as input power and gain spectra, while overlooking internal telemetry that is readily available in modern commercial \acp{EDFA} and could enhance generalisation.}

In our prior work~\cite{fnfw_paper_self}, we introduced a semi-supervised self-normalizing neural network (SS-NN) framework that integrates internal \ac{EDFA} features for gain spectrum modeling, enabling transfer learning across in-built Booster and Preamp \acp{EDFA} in \acp{ROADM}. Building upon this foundation, the present work extends our previous contribution in several key ways:

\begin{itemize}
    \item We extend our experimental validation by collecting gain spectrum measurements on 4 \acp{ILA} (without embedded \acp{OCM}), and investigate the performance of SS-NN model on different EDFA types. 

    \item We investigate whether transfer learning can be applied across different EDFA types (such as Boosters/Preamps), as well as different manufacturers (such as Lumentum in- ROADM EDFAs/Juniper ILAs). 

    \item We introduce a heterogeneous TL method that utilizes covariance matching (CORAL loss) to align the feature representations between source and target domains, thereby addressing the challenges associated with feature mismatches across diverse \ac{EDFA} types and vendors.
    
\end{itemize}

The contributions of this work can be summarized as follows:
\begin{enumerate}
    \item We develop a novel SS-NN architecture that integrates both external and internal \ac{EDFA} features. By combining unsupervised pre-training with supervised fine-tuning, our approach significantly reduces the number of measurements required for effective training.
    \item We introduce a few-shot transfer learning mechanism that enables efficient knowledge transfer from one EDFA to another.
    \item We incorporate a covariance matching technique (CORAL loss) for heterogeneous transfer learning, enhancing domain adaptability across EDFAs with differing feature sets.
    \item We perform a comprehensive experimental evaluation on 26 commercial-grade EDFAs—including Boosters, Pre-Amplifiers, and In-Line Amplifiers (ILAs) from the COSMOS and Open Ireland testbeds—demonstrating improved model performance and reduced measurement requirements.
\end{enumerate}

The remainder of this paper is organized as follows: Section~\ref{section2} describes the experimental setup and data collection methodology for \ac{EDFA} gain spectrum measurements across COSMOS and Open Ireland testbeds. Section~\ref{section3} describes the proposed SS-NN model architecture and training process, and the results on directly-trained models. Section~\ref{section4} discusses the homogeneous and heterogeneous TL methods and results for model transfer across different \ac{EDFA} types. Finally, Section~\ref{section5} summarizes our findings. 

\section{Experimental Setup and Data Collection}
\label{section2}

This section outlines the experimental setup and data collection methodology for \acp{EDFA} in the Open Ireland and PAWR COSMOS testbeds. The Open Ireland testbed~\cite{open_ireland} is an open re-configurable optical-wireless testbed in Dublin, Ireland; while the PAWR COSMOS testbed~\cite{chenSoftwareDefinedProgrammableTestbed2022, raychaudhuri2020challenge} is a city-scale programmable testbed deployed in Manhattan, USA. 

\subsection{Experimental Setup for Booster/Preamp}

\begin{figure}[]
    \centering
    \includegraphics[width=0.8\linewidth]{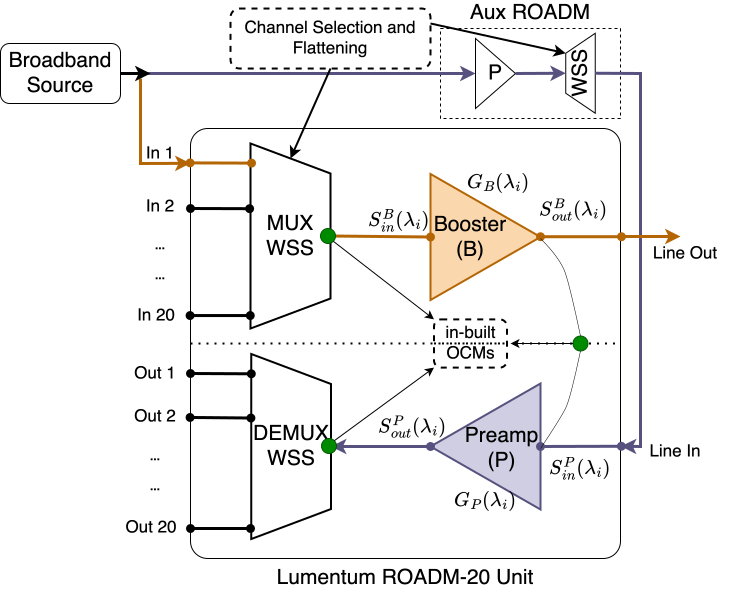}
    \caption{Experimental setup for characterization of Booster/Pre-amplifier \ac{EDFA} of Lumentum \acp{ROADM} in the COSMOS and Open Ireland testbeds.}
    \label{roadm_setup}

\end{figure}

\begin{figure}

    \centering
    \includegraphics[width=\linewidth]{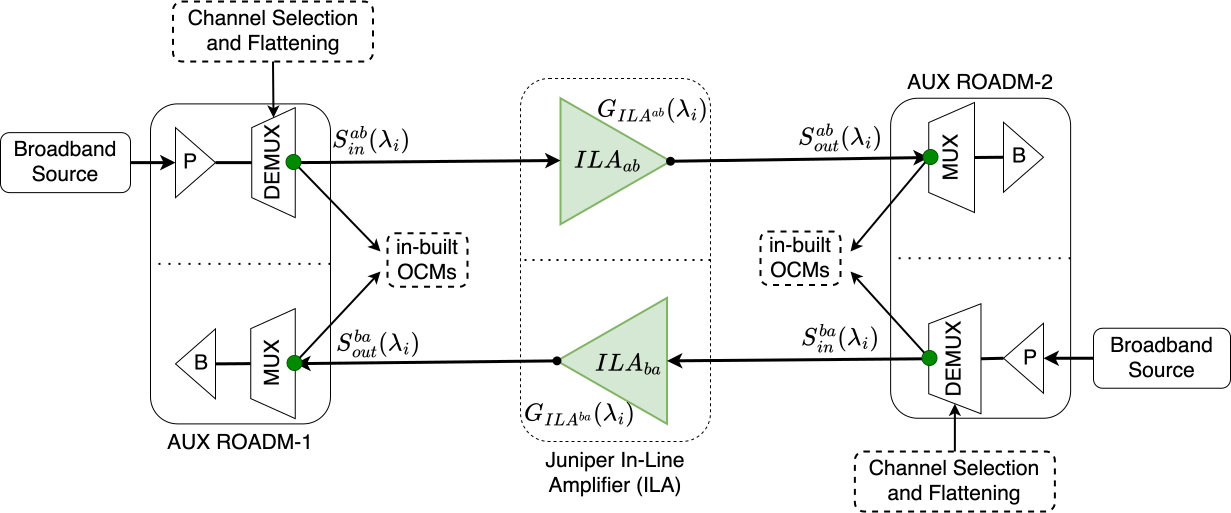}
    \caption{Experimental setup for characterization of the Juniper TCX-1000 \acp{ILA} in the Open Ireland testbed}
    \label{ila_setup}

\end{figure}

Gain spectrum measurements were conducted across multiple C-band wavelengths using commercial-grade Lumentum ROADM-20 units—three deployed in the Open Ireland testbed and eight in PAWR COSMOS. Each unit contains two \acp{EDFA}, yielding data from 11 Booster and 11 Pre-Amplifier \acp{EDFA}. Figure~\ref{roadm_setup} illustrates the experimental topology. A broadband source generates 95 × 50 GHz \ac{WDM} channels in the C-band following the \ac{ITU} \ac{DWDM} 50 GHz grid specification. Data collection procedures were standardized across both testbeds~\cite{wangOpenEDFAGain2023}.

For Booster measurements, the MUX \ac{WSS} flattens channels and controls power and loading configurations. In Pre-Amplifier measurements, the comb source connects to an auxiliary \ac{ROADM}’s Line-IN port, with its DEMUX managing power and channel loading before forwarding the signal to the \ac{EDFA} under test. Input and output power spectra for all 95 channels are recorded via built-in \acp{OCM}, while total input/output power through the \acp{EDFA} is captured via integrated \acp{PM}. Additionally, we collect the internal \ac{VOA} input/output power ($P^V_{in}$, $P^V_{out}$) and attenuation ($P^V_{attn}$) measurements. 

\subsection{Experimental Setup for ILA}

\begin{figure}[]
    \centering
\includegraphics[width=\linewidth]{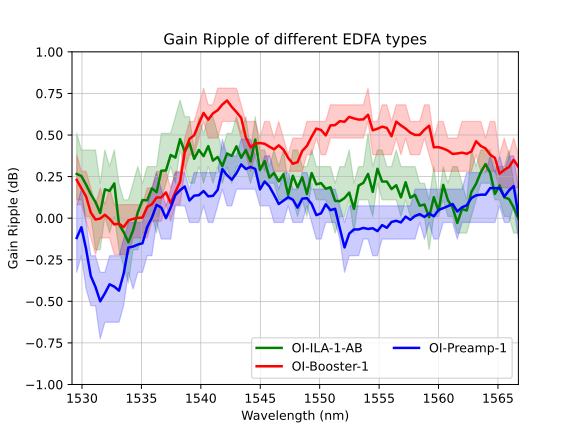}
    \caption{Measured gain ripple for different \ac{EDFA} types (\ac{ILA}, Booster and Preamp) at the same target gain setting of 15 dB. One device of each \ac{EDFA} type used in the OpenIreland Testbed is shown. Solid lines show the mean
gain ripple, while the shaded areas represent the min/max range of gain ripple values.}
    \label{different_edfa_types}
\end{figure}

\begin{figure}[]
    \centering
\includegraphics[width=0.8\linewidth]{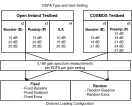}
    \caption{Complete set of measurements collected from \acp{EDFA} across COSMOS and Open Ireland testbeds.}
    \label{measurement_setup}

\end{figure}

We collect gain spectrum measurements from 2 commercial grade, bidirectional Juniper TCX-1000 ILA units deployed in the Open Ireland testbed. The data is collected for both forward and backward directions (typically denoted as AB and BA). Although housed on the same equipment, the forward and backward direction amplifiers function independently. 

Figure~\ref{ila_setup} shows the experimental topology, as in previous setups for boosters/preamps, a broadband source is used to generate 95 x 50 GHz \ac{WDM} channels in the C-band. Unlike \acp{ROADM}, \acp{ILA} do not have a built-in \ac{OCM} which makes data collection difficult. Therefore, we employ two auxiliary \acp{ROADM} to collect gain spectrum measurements from both \acp{EDFA} in a single \ac{ILA}. For the forward direction~(AB), the input and output power spectra are collected using built-in \ac{OCM} at the DEMUX output of first Auxiliary \ac{ROADM}, and at the MUX input of the second auxiliary ROADM; and vice versa for the backward direction EDFA~(BA). The total input/output power through the \acp{ILA} are collected through built-in \ac{PM} in the \acp{ILA}. It should be noted that there will be insertion loss between the actual input/output power spectra between Auxiliary \acp{ROADM} and input/output of the \acp{ILA}. To compensate for these losses, a scaling factor~(\(\sigma\)) is applied to reflect the actual input/output power spectrum values at \ac{ILA} input/output respectively. For each measurement, the scaling factor~(\(\sigma\)) is calculated as follows:

\begin{equation}
    \sigma_{in} = P_{out}^{Aux}/P_{in}^{ILA}
    \label{eq:1}
\end{equation}
and
\begin{equation}
    \sigma_{out} = P_{out}^{ILA}/P_{in}^{Aux}
    \label{eq:2}
\end{equation}
Where \(P_{in}^{Aux}\) and \(P_{out}^{Aux}\) are the total input and output powers of the input and output \acp{ROADM} in milliwatts~(mW), respectively. \(P_{in}^{ILA}\) and \(P_{out}^{ILA}\) are the total input and output powers of the \ac{ILA} \ac{EDFA} under test. The total power readings are measured through the in-built \acp{PM} in both \acp{ROADM} and \acp{ILA}. From \ref{eq:1} and \ref{eq:2}, the normalized power readings for each channel~\(i\) is calculated as follows: 

\begin{equation}
    P_{in}(\lambda_i) = \sigma_{in} \cdot P_{in}^{Aux}(\lambda_i)
\end{equation}
and 
\begin{equation}
    P_{out}(\lambda_i) = \sigma_{out} \cdot P_{out}^{Aux}(\lambda_i)
\end{equation}
where \(P_{in}^{Aux}(\lambda_i)\) and \(P_{out}^{Aux}(\lambda_i)\) are the measured \ac{OCM} power readings at the Auxiliary \acp{ROADM}, for each channel \(i\) in the 95 X 50 GHz \ac{DWDM} channels. It should be noted that unlike Boosters/Pre-amplifiers, the internal \ac{VOA} measurements are not exposed for \acp{ILA}. 
\begin{figure*}[]
    \centering

    \includegraphics[width=0.32\linewidth]{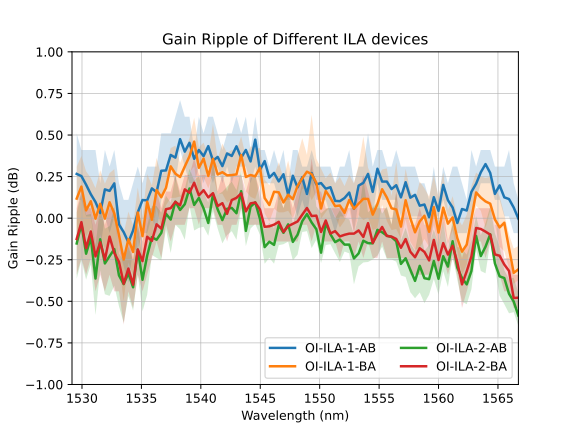}    \includegraphics[width=0.32\linewidth]{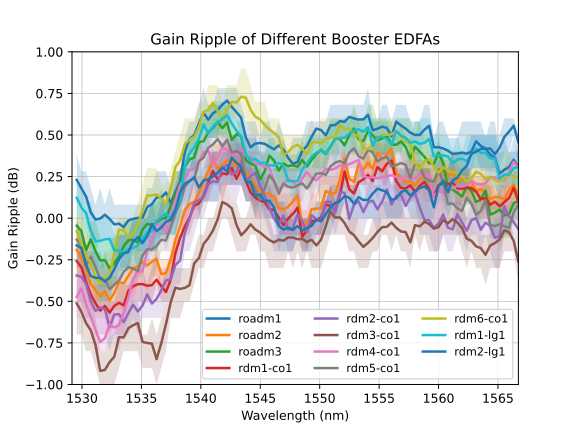}
    \includegraphics[width=0.32\linewidth]{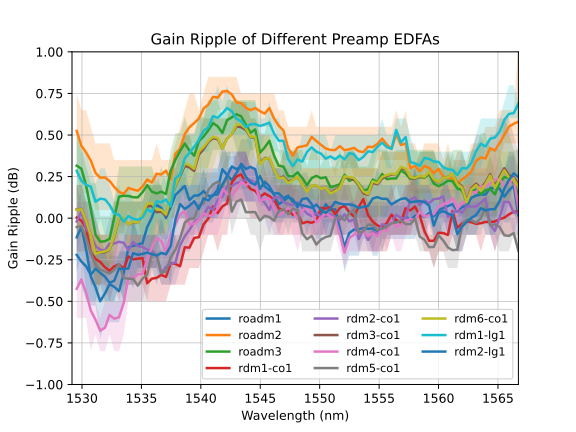}
    \caption{Measured gain ripple of same \ac{EDFA} type, across different devices at the same target gain setting. Solid lines show the mean gain ripple, while the shaded areas represent the min/max range of gain ripple values.}
    \label{booster_gain_ripple}

\end{figure*}

\subsection{Measurement Configuration}

In the Open Ireland testbed, all Booster and Preamp \acp{EDFA} were characterized at target gain settings of 15/20/25 dB, while in the COSMOS testbed, the target gains were set to 15/18/21 dB for Boosters and 15/18/21/24/27 dB for Pre-Amplifiers in high gain mode. All \acp{ILA} were measured in the Open Ireland testbed at target gains of 10/15/20 dB in low gain mode. All measurements were taken with 0 dB gain tilt. The use of varying gain configurations across different \ac{EDFA} types and testbeds is intended to replicate the operational diversity encountered in different network environments. The dataset comprises 3,168 gain measurements (collected across multiple wavelengths) per \ac{EDFA} for each designated target gain setting. Overall, the COSMOS testbed yielded 202,752 measurements across 11 boosters and 11 preamps. In the Open Ireland testbed, a total of 57,024 measurements were taken in 3 boosters and 3 preamps, while 38,016 measurements were taken from 4 \acp{EDFA} in 2\acp{ILA}~(in both forward and backward directions). All measurements are collected under three different channel loading configurations: 
\begin{enumerate}
    \item \textit{Fixed}: This includes fully loaded (WDM), half-loaded (lower/upper spectrum, and even/odd numbered channels), adjacent single/double channel loads, as well as a complete set of single/double channel configurations. 
    \item \textit{Random Allocation}: This configuration comprises of random channel configurations from small-scale single-channel loads to fully loaded setups. 
    \item \textit{Goalpost Allocation}: This includes structured channel-loading configurations across different wavelength bands (short, medium and long wavelengths); with both balanced and unbalanced loads across the considered bands. 
\end{enumerate}

Fig.~\ref{different_edfa_types} shows the measured gain ripple at a target gain setting of 15 dB for a Booster, Pre-amplifier and an \ac{ILA} in the Open Ireland testbed. It can be seen that different \ac{EDFA} types exhibit different gain ripples. Fig.~\ref{booster_gain_ripple} shows the gain ripples of different devices measured across Booster, Pre-amplifier and \ac{ILA}. It can be seen that even devices from the same manufacturer exhibit different gain ripple at the same target gain setting.

    





\section{Model Architecture}
\label{section3}

\begin{figure}[]
    \centering
    \includegraphics[width=0.8\linewidth]{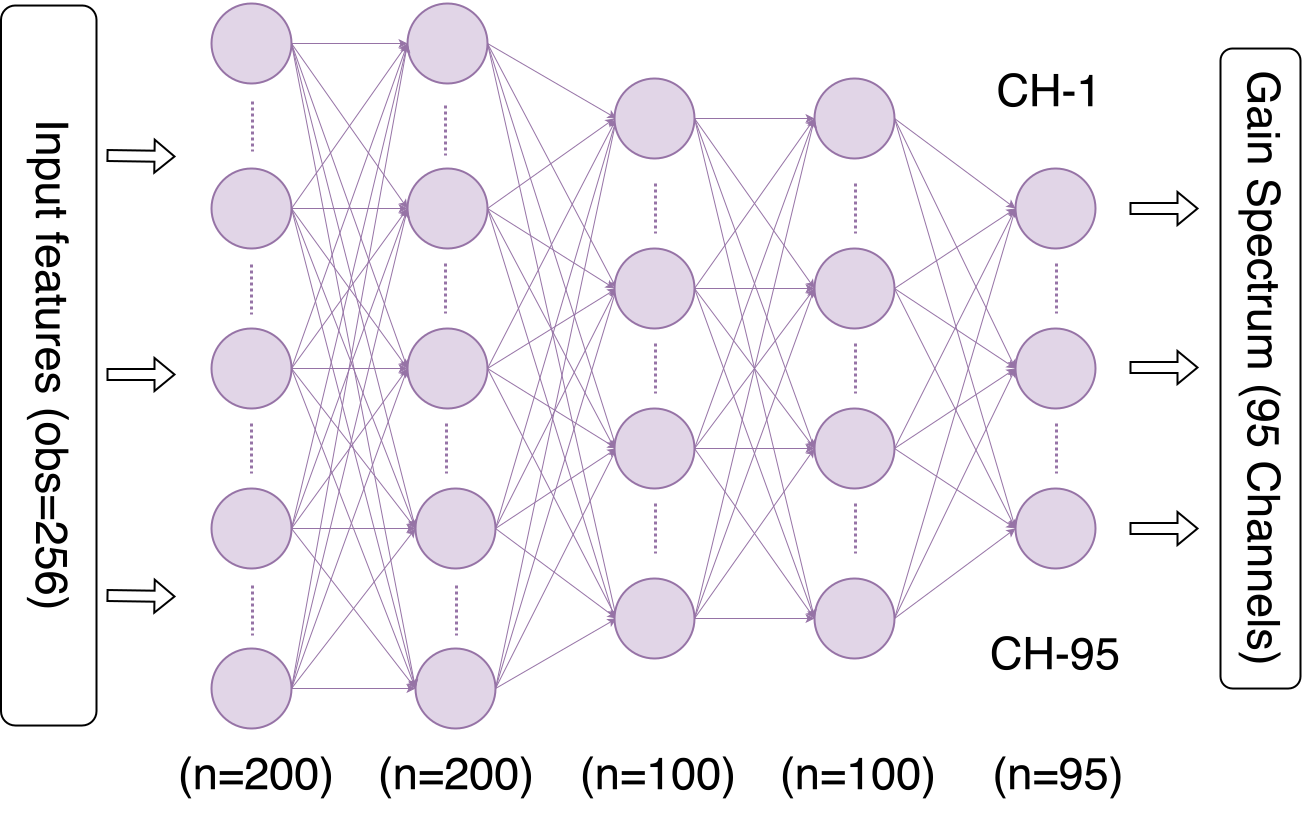}
    \caption{SS-NN model structure with 5 layers.}
    \label{model_structure}
\end{figure}

The Semi-Supervised Self-Normalizing Neural Network~(SS-NN) model is designed to address two challenges in EDFA gain spectrum characterization: limited training data and poor transferability across different EDFA devices. The proposed model integrates internal features of the EDFA, self-normalizing activation function, and a two-stage training process, which enables few-shot learning and improved generalization capabilities. In the following sections, we provide a detailed explanation of the model architecture and the training methodology.

\subsection{SS-NN Model}

Existing ML approaches for EDFA modeling often treat the amplifier as a blackbox, replying solely on input/output spectra. For boosters and preamps, we incorporate three key features derived from the internal \ac{VOA}: input/output power ($P^V_{in}$, $P^V_{out}$) and attenuation ($P^V_{attn}$). \acp{VOA} are an internal component of \acp{EDFA}, which indirectly influence the shape of the gain profile by acting on the signal's input powers. This is done to ensure the \ac{EDFA} operates in its design average inversion for a flat spectrum gain profile which matches the \ac{GFF} attenuation~\cite{zyskind_optically_2011}. The \ac{VOA} attenuation is controlled automatically in the \ac{EDFA} based on the model's gain dynamic range, and it grants intrinsic information on the operation of each \ac{EDFA}. In addition to the VOA parameters, the full input feature set includes the target gain ($G_0$) with constant-gain configuration, total EDFA input/output power ($P_{in}$, $P_{out}$), input power spectrum ($\overline{P}{(\lambda_i)} = [P{(\lambda_1)}, P{(\lambda_2)}, P(\lambda_3), ... P{(\lambda_{95})}]$), and a binary channel-loading vector $\overline{C}=[c_i]_{i=1}^{95}$, where
\begin{equation}
    c_i = \begin{cases}
    1, & \text{if the \(i^{th}\) wavelength channel is active} \\
    0, & \text{otherwise.}
    \end{cases}
\end{equation}

Figure~\ref{model_structure} shows the overall architecture of the SS-NN model utilizes a four hidden layer topology with 200/200/100/100 neurons. The input layer consists of 196 neurons, while the output layer consists of 95 neurons in all cases, predicting the gain spectrum at 95 wavelengths,  $G(\lambda_i) = [G(\lambda_1), \cdots, G(\lambda_{95})]$. \added{The final model architecture was determined through a random search over the subspace of (i) number of hidden layers $\in \{1, 2, 4, 8, 10\}$, and (ii) number of neurons in each hidden layer $\in \{50, 95, 100, 150, 200, 300\}$. The final model architecture with 4 hidden layers and a neuron count of (200-200-100-100) was chosen optimally on the basis of model accuracy as well as the pre-training computational cost.} The model is based on \acp{SNN} with \ac{SELU} as the activation function, where the \ac{SELU} function is defined as:

\begin{equation}
\text{selu}(x) = \lambda x \cdot I(x>0) + \alpha (e^x-1)\cdot I(x\leq 0)
\end{equation}
where $\alpha = 1.673$ and $\lambda = 1.050$, and ${I}$ denotes the indicator function. In \acp{ILA}, the missing internal \ac{VOA} features are imputed with an extreme value of -999, in order to saturate the output of input neurons corresponding to the missing input features. In neurons with \ac{SELU} activation function, the gradient becomes saturated at highly negative inputs such as -999, essentially acting as a dead neuron. 

Typically, batch normalization is commonly applied to stabilize training by normalizing the outputs of hidden layers~\cite{Wang:23}. However, batch normalization tends to perform poorly when training with limited data, as it relies on the estimation of batch statistics that can be unreliable in such scenarios~\cite{NIPS2017_c54e7837}. In contrast, the \acp{SNN} with \ac{SELU}~\cite{klambauerSelfNormalizingNeuralNetworks2017a} activation function ensures stable variance propagation across layers, even with less data. This self-normalizing property is a key factor enabling the SS-NN architecture to effectively perform one-shot training and achieve cross-device transferability.

\subsection{Training Process}

\begin{figure*}[]
    \centering
    \includegraphics[width=\linewidth]{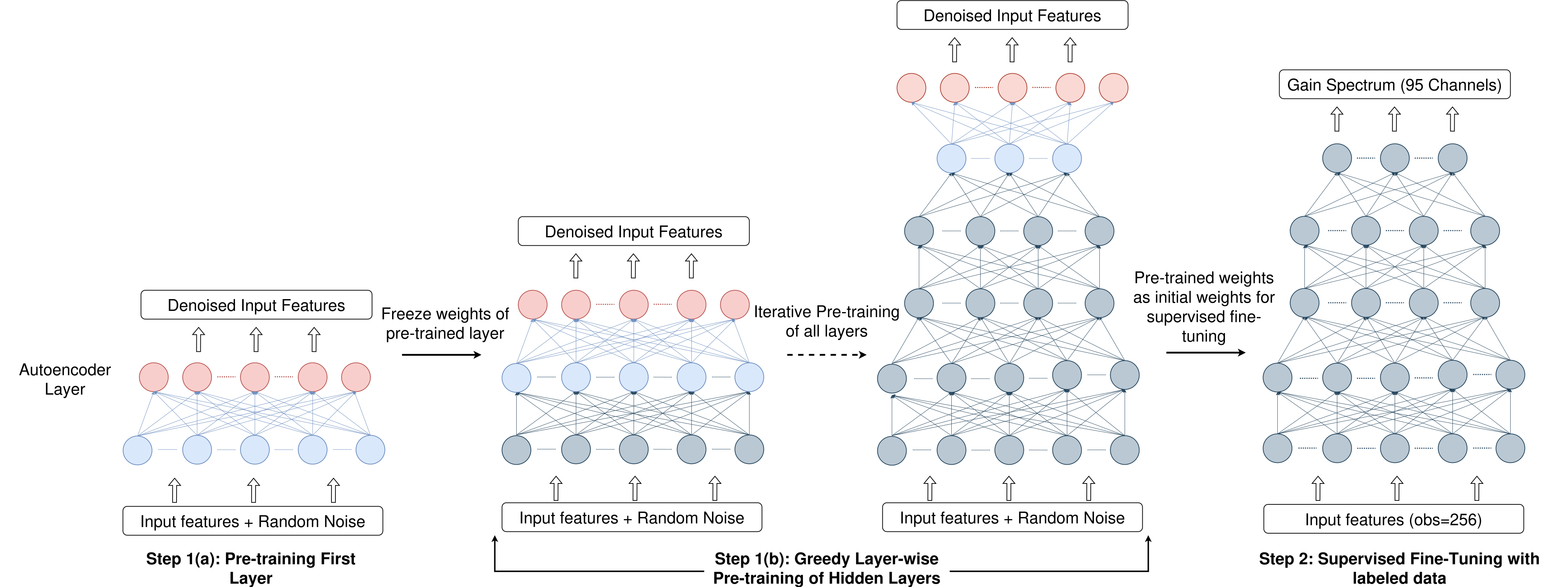}
    \caption{SS-NN model training framework. Step 1(a) and (b) show the greedy layer-wise pretraining of hidden layers using unsupervised pretraining. This pre-trained model forms the basis for Step 2, where supervised fine-tuning is performed with 256 labeled measurements.}
    \label{base_model}
\end{figure*}

The proposed model employs a two-phase training methodology comprising unsupervised pre-training~\cite{NIPS2006_5da713a6, geProvableAdvantageUnsupervised2023} followed by supervised fine-tuning~\cite{s22114157}. This approach is driven by two considerations:
\begin{enumerate}
    \item Unsupervised pre-training leverages readily available unlabeled input power spectrum measurements. These data points are more accessible and can also be simulated in flat spectrum scenarios, which significantly mitigates the need for resource-intensive labeled datasets.
    \item Pre-training process helps achieve a better weight initialization compared to random initialization and captures more complex dependencies between parameters~\cite{geProvableAdvantageUnsupervised2023}. Furthermore, it introduces implicit regularization properties that improve generalization, particularly beneficial when transferring knowledge between different \ac{EDFA} configurations and units~\cite{erhan2010does}.
\end{enumerate}

Figure~\ref{base_model} illustrates the training procedure. During unsupervised pre-training, weights are initialized progressively using a layer-wise denoising strategy, where each layer is trained on 512 noise-augmented measurements per gain setting. Gaussian noise is added to the inputs, and an autoencoder structure, which matches the dimensionality of the original feature set, is tasked with reconstructing denoised outputs. The training employs \ac{MSE} loss to evaluate reconstruction fidelity under noise corruption. Each layer undergoes sequential training for 1,800 epochs (learning rate = 1e-03), with its weights frozen upon initialization to preserve hierarchical feature representations during subsequent training stages. Once all layers are pre-trained, the finalized model serves as the foundation for supervised fine-tuning.

For supervised fine-tuning, the model is optimized using a labeled dataset of 512 measurements containing fully loaded and random channel configurations. We utilizes a custom weighted \ac{MSE} loss function, for \(k^{th}\) measurement it is defined as:

\begin{equation}
\label{loss_function}
\text{MSE}_k = \frac{1}{\sum_{i=1}^{95} c_i^k} \cdot \sum_{i=1}^{95} c_i^k . \left[ g_{\text{pred}}^k(\lambda_i) - g_{\text{meas}}^k(\lambda_i) \right]^2
\end{equation}
In this phase, the model is fine-tuned using the Adam optimizer with a learning rate of 1e-03, over 1,200 epochs \added{with a batch size of 32}. \added{To account for the wide dynamic range of gain‑spectrum measurements obtained across different power levels and channel configurations, we apply gradient clipping~\cite{Pascanu2012-up} with a threshold of 1.0 to stabilize training and prevent divergence.} \added{During the entire training process, hyperparameter optimization was performed using random search to fine-tune learning rate, number of layers, number of neurons in each layer and batch size.}

\subsection{Training and Test Sets}

We compare the SS-NN model with a benchmark state-of-the-art method~\cite{Wang:23, wangOpenEDFAGain2023}. For equivalent comparison, we follow the same dataset selection criteria. For each gain setting, we split the dataset into a training/test set ratio of 0.86/0.14. The test set contains 436 gain spectrum measurements per gain setting. This test set contains a mixture of random and goalpost channel loading measurements, which represent a diverse set of channel loading configurations. Note that although the SS-NN model uses less data for training, we allocate a larger portion of training data for the benchmark model, which uses 2,732 measurements per gain setting. 

\subsection{Model Performance}

We compare the SS-NN model with the benchmark model using the same set of features to highlight the benefits of our approach. Additionally, for boosters/preamps, we demonstrate the advantage of incorporating internal \ac{EDFA} features by comparing the results of the SS-NN model with and without including these additional features. 

\begin{figure}
    \includegraphics[width=0.92\linewidth]{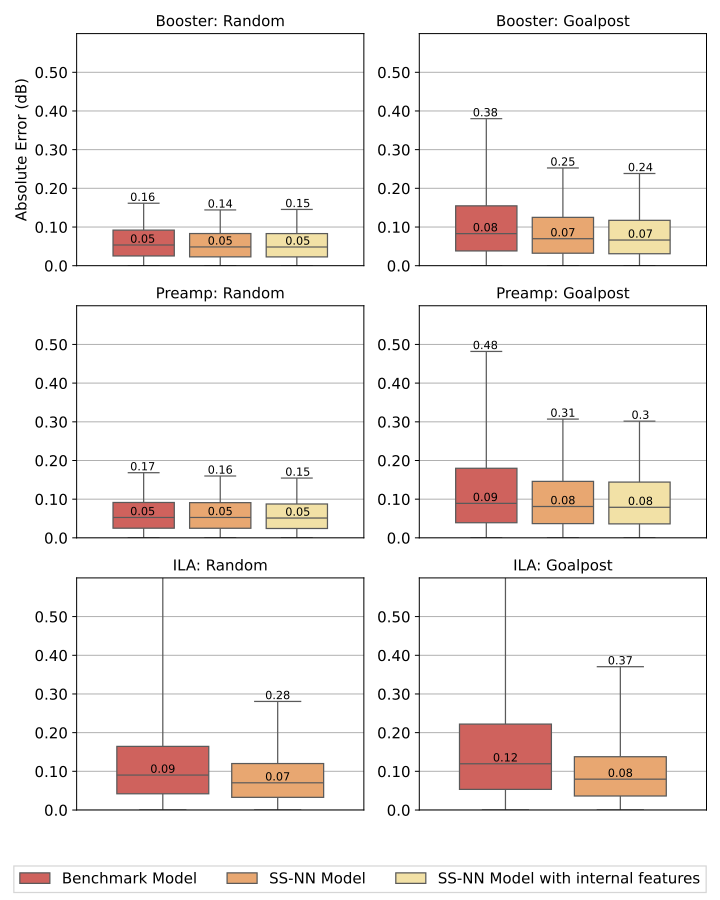}
    \caption{Boxplot distribution of absolute errors across all 11 Boosters, 11 Pre-amplifiers and 4 \ac{ILA} EDFAs for goalpost and random channel loading. The boxes denote the inter-quartile range, and the whiskers denote the min/95th percentile. }
    \label{boxplot}
\end{figure}

Figure~\ref{boxplot} shows the distribution of absolute errors of gain spectrum predicted by the benchmark model, SS-NN model using same set of features, and SS-NN model with additional internal \ac{VOA} features. The errors are calculated across 11 boosters, 11 pre-amplifiers and 4 \ac{ILA} \acp{EDFA} in the OpenIreland and COSMOS testbeds on the test set with random and goalpost channel configurations. For boosters, the SS-NN model achieves a mean absolute error of 0.05 dB and 0.07 dB under the random and goalpost channel configurations. This is comparable to the performance of the benchmark model which uses a considerably higher number of measurements (8196 measurements), compared to a total of 1,792 measurements utilized by the SS-NN model. Importantly, the SS-NN models exhibit a superior error distribution, with a narrow inter-quartile range, and a 95\textsuperscript{th} percentile error of 0.15/0.24 dB, compared to 0.38/0.16 dB by the benchmark model, across the goalpost/random test sets. 

For preamps, the SS-NN model achieves a mean absolute error of 0.08/0.05 dB using the same set of features, and 0.07/0.05 dB using additional internal features across goalpost/random channel configurations. This is marginally better than the benchmark model which achieves a 0.09/0.05 dB error across goalpost/random test sets. Additionally, the distribution of errors for SS-NN models are more stable, with a narrow inter-quartile range, and a 95\textsuperscript{th} percentile error within 0.3 dB across both channel configurations, showing that the SS-NN model generalizes well to unseen channel configurations even when trained with reduced measurements. It should be noted that using additional internal features when directly training \ac{EDFA} models i.e., training on the source \ac{EDFA}'s measurements without \ac{TL} does not provide additional performance. 

For \acp{ILA}, we only develop the SS-NN model against the Benchmark model, due to non-availability of internal \ac{VOA} measurements in Juniper \acp{ILA}. Moreover, because of measurement setup involving external \acp{ROADM}, the \ac{ILA} input and output spectra exhibit more noise compared to booster/preamps - which are in-\ac{ROADM} \acp{EDFA}. In this case, the SS-NN model performs much better than the benchmark model both in terms of \ac{MAE} and distribution. The model achieves 0.07/0.08 dB error on goalpost datasets, with a 95\textsuperscript{th} percentile error within 0.4 dB across both the channel configurations. 

\begin{table}
    \centering
    \begin{tabular}{|c|c|} \hline 
         \textbf{Parameter}& \textbf{Value}\\ \hline 
         Trainable Parameters& 119, 395\\ \hline 
         Batch Size& 32 \\ \hline
         Total Training Time& 1,180 s\\ \hline
         Inference Latency& 1.1 ms\\ \hline 
         Inference \acp{FLOP}& 238,095\\ \hline 
         
    \end{tabular}
    \caption{\added{Computational Overhead for SS-NN model with largest number of features (including internal features). Latency times are averaged over the entire dataset, and approximated to first decimal place.}}
    \label{tab:latency}
\end{table}

\added{Table~\ref{tab:latency} summarizes the computational overhead for the SS-NN model, including the total number of trainable parameters, per-inference \acp{FLOP} and approximate training times. All experiments were conducted on an Nvidia RTX 4090 GPU in the Open Ireland testbed, with 24 GB of VRAM and a peak 32 bit floating point throughput of 82.58 \ac{TFLOPS}~\cite{UnknownUnknown-an}. Training latency is substantially higher than the inference latency because of a smaller batch size of 32. This was driven by our finding that lower batch size improved convergence when training with a limited set of measurements in the fine-tuning step. The inference time per observation is $\approx$ 1.1 ms, which is considerably faster than the 6 seconds temporal resolution of in-device \ac{OCM} measurements on Lumentum ROADM-20 module~\cite{wangOpenEDFAGain2023}.}

\section{Transfer Learning}
\label{section4}
\begin{figure*}[]

    \centering
    \includegraphics[width=\linewidth]{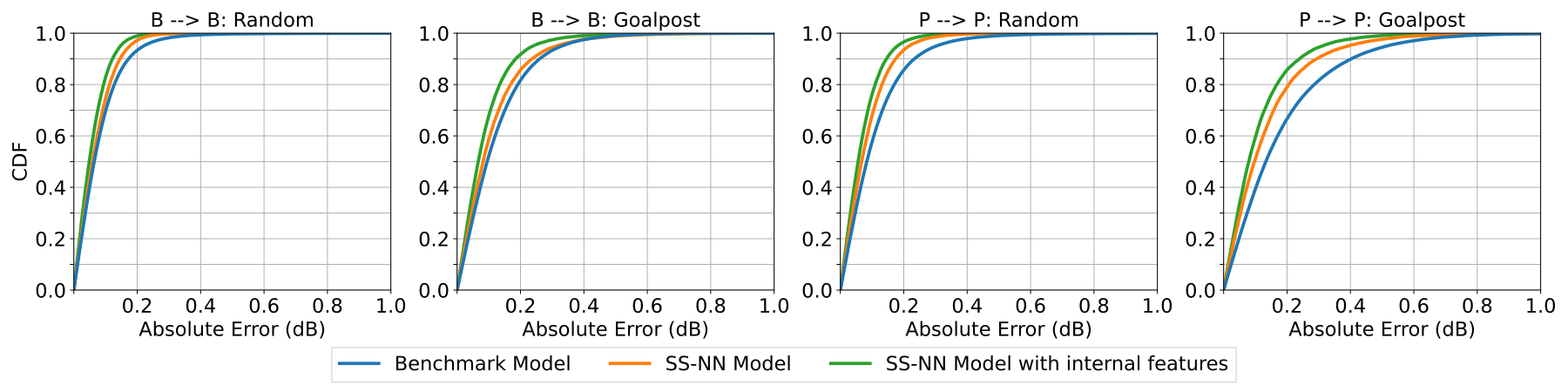}
    \caption{Cumulative Distribution Function~(CDF) plots of absolute errors for TL across same-type homogeneous TL for (i) Booster to Booster TL: Random channel loading, (ii) Booster to Booster TL: Goalpost channel loading, (iii) Preamp to Preamp TL: Random channel loading, and (iv) Preamp to Preamp TL: Goalpost channel loading.}
    \label{same_type_cdf}
\end{figure*}

\begin{figure}
    \centering
\includegraphics[width=\linewidth]{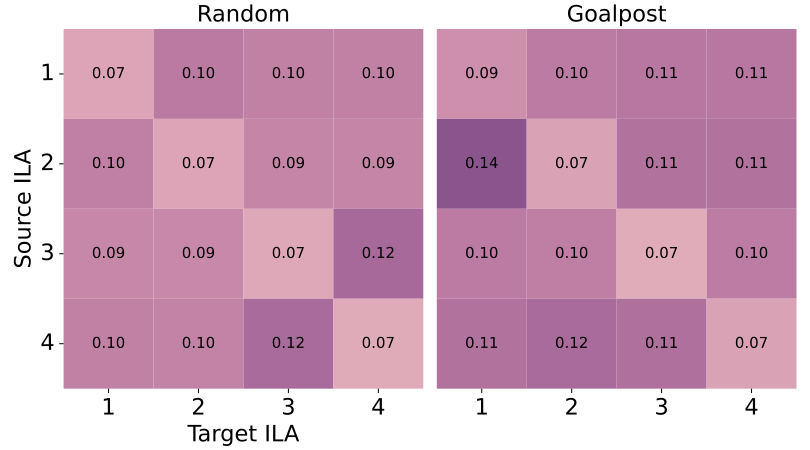}
    \caption{Transfer Learning MAE matrix of SS-NN model for 4 \acp{ILA} in the Open Ireland testbed, shown across random and goalpost channel loading configurations. The (i, j) entry corresponds to the TL-based \ac{ILA} model, where the $i^{th}$ and $j^{th}$ EDFA serve as the source and target models, respectively.}
    \label{ila_heatmap}

    \includegraphics[width=\linewidth]{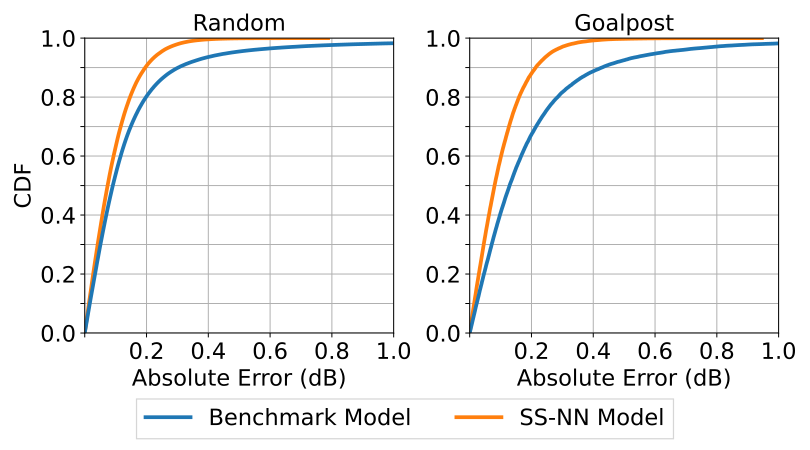}
    \caption{Cumulative Distribution Function (CDF) plots of absolute errors for TL across \acp{ILA} in the Open Ireland testbed.}
    \label{ila_cdf}
\end{figure}

\begin{figure*}[]
    \centering
    \includegraphics[width=\linewidth]{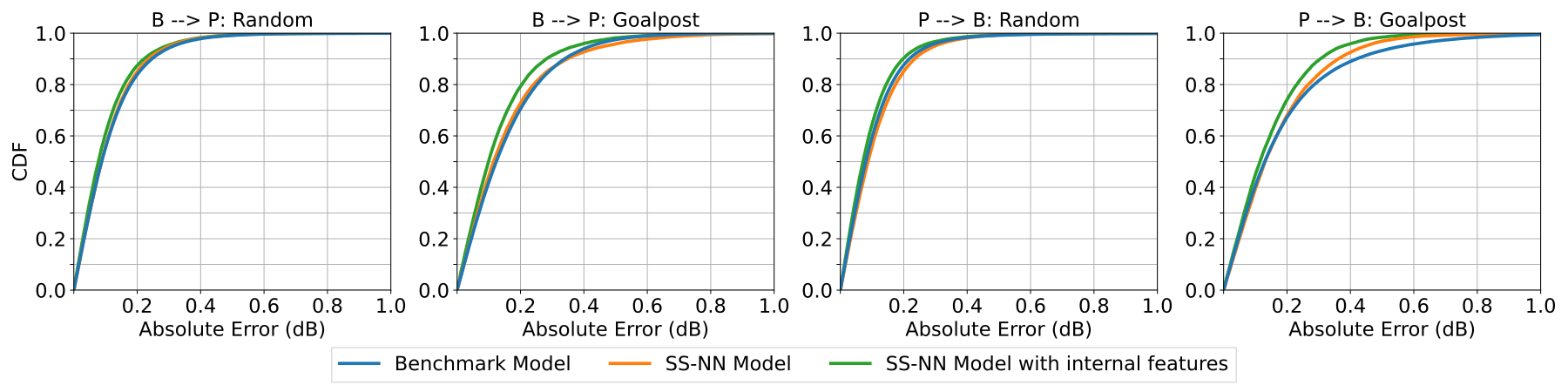}

    \caption{Cumulative Distribution Function~(CDF) plots of absolute errors for TL across cross-type homogeneous TL for (i) Booster to Preamp TL: Random channel loading, (ii) Booster to Preamp TL: Goalpost channel loading, (iii) Preamp to Booster TL: Random channel loading, and (iv) Preamp to Booster TL: Goalpost channel loading.}
    \label{cross_type_cdf}

\end{figure*}
Transfer Learning (TL) improves performance on a target task by leveraging information from a related source domain~\cite{zhuangComprehensiveSurveyTransfer2021a}. In the context of \ac{ML} algorithms, TL involves fine-tuning a model trained on a source domain (\(D_S\)) using new feature representations from a target domain (\(D_T\)). This approach is particularly useful for modeling the gain spectrum in \acp{EDFA}, as it can reduce the required measurement time.

For a source domain \(D_S\) with \(n\) features, the feature representation for an instance \(i\) can be expressed as:
\[
X_s^i = \{x_{s1}^i, x_{s2}^i, \dots, x_{sn}^i\}.
\]
Similarly, for a target domain \(D_T\) with \(m\) features, an instance \(j\) is represented by:
\[
X_t^j = \{x_{t1}^j, x_{t2}^j, \dots, x_{tm}^j\}.
\]
Given a model \(f_s\) trained on \(D_S\), TL is performed by augmenting \(f_s\) with additional instances from \(D_T\). When \(n = m\), TL is considered \emph{homogeneous}, whereas when \(n \neq m\), it is classified as \emph{heterogeneous}. In the latter case, differences in feature space dimensionality can lead to information loss and degraded model performance. Furthermore, if the source domain contains more information than the target (\(n > m\)), negative transfer may occur~\cite{weissSurveyTransferLearning2016, oroujiDomainAdaptationSmallscale}, where the source knowledge adversely affects the target model. Although training both models on only shared features could improve performance, it would compromise the source model’s accuracy.

In the case of \ac{EDFA} TL, a similar trade-off is evident. As discussed previously, incorporating additional internal \ac{EDFA} features enhances model performance. However, a source model trained with these extra features cannot be directly transferred to a target \ac{EDFA} (e.g., an \ac{ILA}) that lacks the ability to measure them. Conversely, training a source model without these features would enable direct transfer but require maintaining dual models or accepting suboptimal performance in the source domain. Based on these considerations, we categorize \ac{EDFA} TL as follows:

\begin{enumerate}
    \item \textbf{Homogeneous TL}: This applies to TL between \acp{EDFA} with identical feature spaces, including:
    \begin{enumerate}
        \item \textit{Same-Type Transfers}: \(B \leftrightarrow B,\; P \leftrightarrow P,\; ILA \leftrightarrow ILA\)
        \item \textit{Cross-Type Transfers}: \(B \leftrightarrow P\)
    \end{enumerate}
    \item \textbf{Heterogeneous TL}: This covers transfers between models trained on Boosters/Pre-amplifiers (with the full feature set) and \acp{ILA} (which lack internal features), i.e., \(B/P \leftrightarrow ILA\).
\end{enumerate}

Note that same-type TL and cross-type TL between Boosters and Preamps are categorized as homogeneous TL, as are transfers among \acp{ILA}, given their consistent feature space. In contrast, TL between an \ac{ILA} and a Booster/Preamp is heterogeneous due to the missing internal \ac{VOA} features in \acp{ILA}.

In this section, we demonstrate that TL for SS-NN models can effectively model the gain spectrum across different \acp{EDFA} with minimal additional data. Section~\ref{homo_tl} details the TL techniques and results for same-type and cross-type transfers, while Section~\ref{hetero_tl} investigates transfers between \acp{EDFA} with differing feature sets and presents an updated TL approach for heterogeneous scenarios.
\subsection{Homogeneous Transfer Learning}
\label{homo_tl}

To transfer an existing model from a source \ac{EDFA} to a target \ac{EDFA}, we fine-tune the source model using a single fully-loaded measurement per target gain setting. The model is fine-tuned using the Adam optimizer for 10,000 epochs, employing the \ac{MSE} loss function as in Eq.~(\ref{loss_function}) and a gradient clipping threshold of 1.0 to ensure stable training. Instead of using a uniform learning rate, a differential learning rate is applied across layers. \added{Rather than completely freezing initial layers, we adopt an exponentially decaying, layer-wise differential learning rate strategy~\cite{Howard2018-gl, Yosinski2014-ch}. The layer-specific learning rate~$\alpha_l$ for layer $l$ in a neural network with total $L$ layers can be given as:

$\alpha_l = \alpha_0\cdot10^{\theta\cdot(L-l)}$, 

where $\alpha_0$ is a constant denoting the base learning rate, and $\theta$ is the exponential multiplier. Empirically, we found the values of $\alpha_0=10^{-3}$ and $\theta=-1$ to work well with convergence.}This strategy enables the output layer to adapt rapidly to the target \ac{EDFA}'s specific characteristics, while the lower layers are fine-tuned more conservatively to prevent overfitting and promote robust generalization.

\subsubsection{Same-type Transfer Learning}

Figure~\ref{same_type_cdf} presents the cumulative distribution function (CDF) of absolute errors for same-type TL among booster/preamp \acp{EDFA} under both random and goalpost channel loading configurations. Error distributions are provided for all possible Booster-to-Booster (B\(\rightarrow\)B) transfers across 11 Booster \acp{EDFA} and Preamp-to-Preamp (P\(\rightarrow\)P) transfers across 11 Preamp \acp{EDFA} from the COSMOS and Open Ireland testbeds. We compare the SS-NN model with (i) the benchmark TL technique~\cite{wangOpenEDFAGain2023}, which employs the same NN architecture, and (ii) the SS-NN model trained without internal \ac{EDFA} features. This comparison highlights the benefit of incorporating additional internal variables along with the proposed modeling technique. Importantly, the benchmark model requires 13 additional measurements per target gain setting for TL, whereas the SS-NN model requires only one.

The error distribution—particularly in the 90th percentile—is critical since \ac{NN} models are prone to overfitting on data subsets. The SS-NN models, however, exhibit a more favorable error distribution for both B\(\rightarrow\)B and P\(\rightarrow\)P transfers under both channel loading configurations. While training directly with internal \ac{EDFA} features does not yield a substantial performance benefit, these variables enhance TL performance, indicating that they contain distinctive information about the \acp{EDFA} under test.

Figure~\ref{ila_cdf} shows the CDF of absolute errors for TL across four \acp{ILA} under random and goalpost channel loading configurations in the Open Ireland testbed. The SS-NN model demonstrates a superior error distribution compared to the benchmark model, particularly in the 95th percentile where the absolute error is \(\leq\) 0.2 dB in both test sets. Figure~\ref{ila_heatmap} displays the MAE matrices (in dB) for the SS-NN model across four \acp{ILA} under both channel loading configurations. In each matrix, the diagonal entry \((i,i)\) corresponds to a directly trained model (without TL), while the off-diagonal entry \((i,j)\) corresponds to a transferred EDFA model with the \(i^{th}\) and \(j^{th}\) \acp{EDFA} serving as the source and target, respectively. These results indicate that the SS-NN model generalizes well to other \acp{ILA} even when TL is performed with only a single measurement per target gain setting. Specifically, the SS-NN-based TL model achieves a per-EDFA MAE below 0.14 dB across both test sets.

\subsubsection{Cross-type Transfer Learning}

Cross-type TL in \acp{EDFA} is more challenging than same-type TL due to significant differences in the feature space. Boosters and preamps are designed for different purposes: preamps are low-noise, high-gain \acp{EDFA} that can substantially improve a transceiver's \ac{SNR}, whereas boosters are optimized for operation in the gain saturation range to enable long-distance transmission~\cite{zyskind_optically_2011}.

Figure~\ref{cross_type_cdf} shows the CDF plots of absolute errors for cross-type TL between boosters and preamps, evaluated under both random and goalpost channel loading configurations. The results include error distributions for all possible combinations of 11 boosters and 11 preamps from the COSMOS and Open Ireland testbeds. As before, comparisons are made with the benchmark TL model and the SS-NN model trained without internal \ac{EDFA} features. Although cross-type TL performance is slightly inferior to same-type TL, the SS-NN model still demonstrates a more favorable error distribution than the benchmark. In particular, under random channel loading, the SS-NN model achieves an absolute error of \(\leq\) 0.2 dB for 95\% of the measurements and outperforms the benchmark in goalpost configurations. Enhancing the error distribution in goalpost scenarios remains an area for future work.

\begin{figure*}

    \centering
    \includegraphics[width=\linewidth]{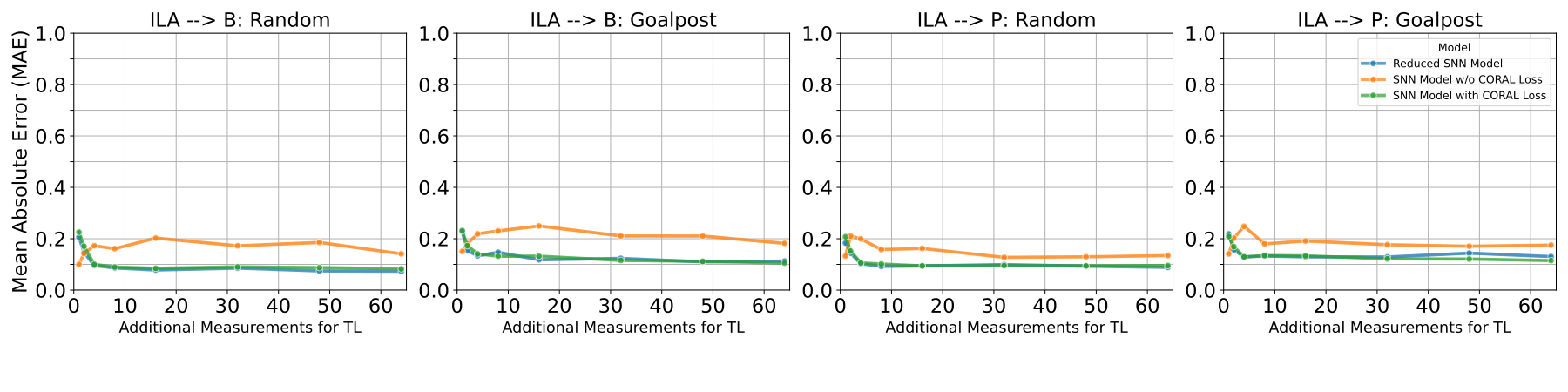}
    \includegraphics[width=\linewidth]{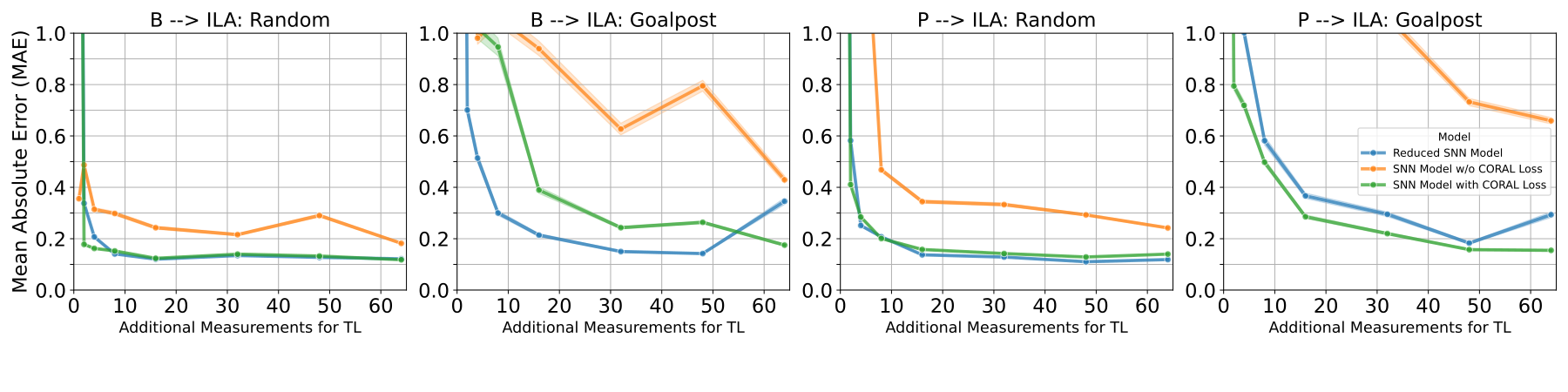}
    \caption{Mean Absolute Error~(MAE) for TL across booster/preamp and ILAs, shown across number of additional measurements per target-gain setting required for TL. Results are shown for (i) Source model trained without internal EDFA features (Reduced SNN Model), (ii) Source model trained with internal EDFA features, but TL without CORAL loss, and (iii) Source model trained with additional features, and TL incorporating CORAL loss.}
    \label{ila_tl_mae}
\end{figure*}

\begin{figure*}

    \centering
    
    \includegraphics[width=\linewidth]{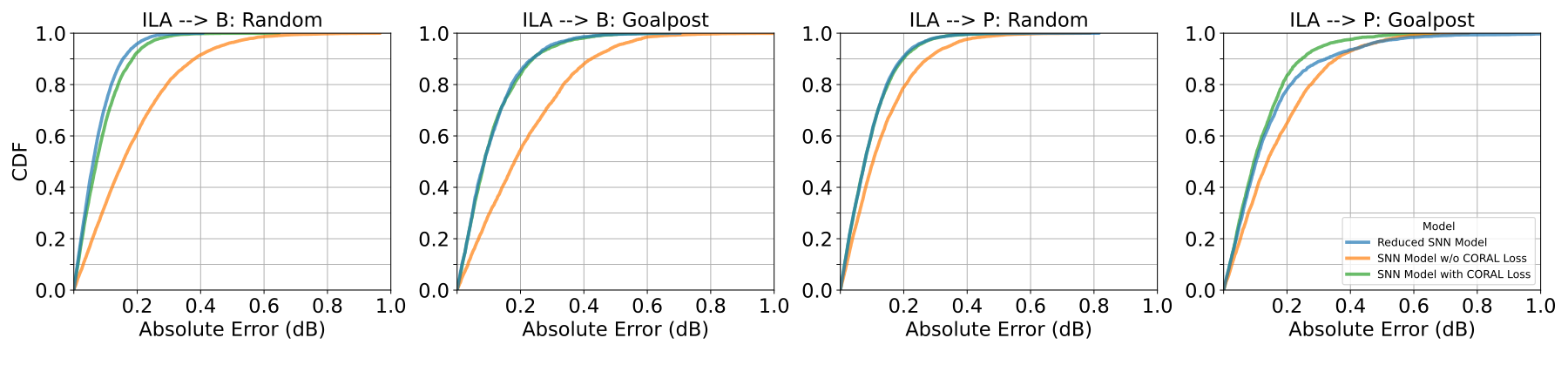}
    \includegraphics[width=\linewidth]{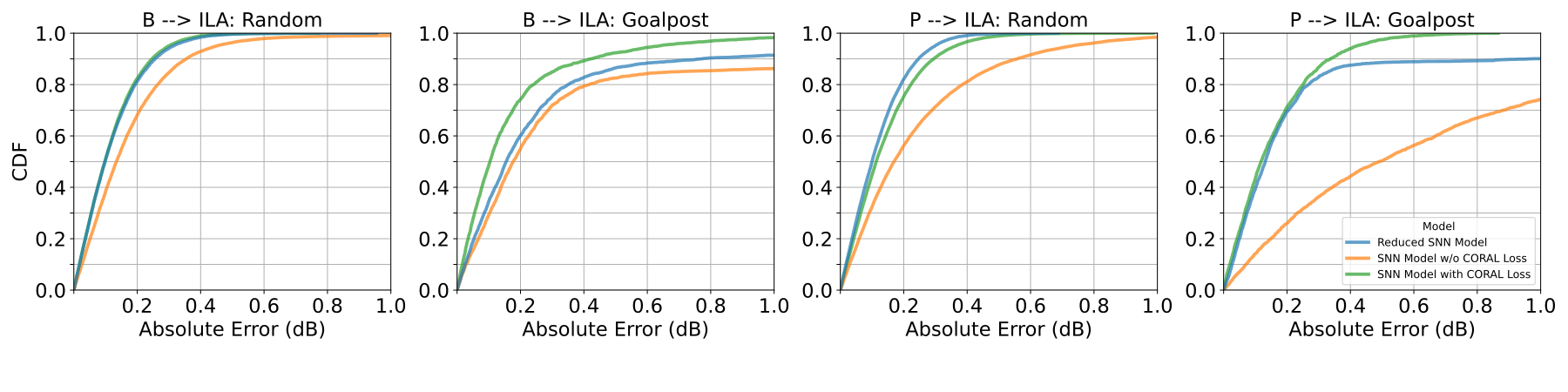}
    \caption{Cumulative Distribution Function~(CDF) Plots for TL across boosters/preamps and ILAs for random and goalpost channel loading configurations. (i) For TL from ILAs\(\rightarrow\)booster/preamp, the target model is fine-tuned with 32 additional measurements per target-gain setting for each target EDFA; while for TL from booster/preamp\(\rightarrow\)ILAs, 48 such additional measurements are required. Results are shown for (i) Source model trained without internal EDFA features (Reduced SNN Model), (ii) Source model trained with additional features, but TL without CORAL loss, and (iii) Source model trained with additional features, and TL incorporating CORAL loss.}
    \label{ila_tl_cdf}
\end{figure*}

\subsection{Heterogeneous Transfer Learning}
\label{hetero_tl}

Transferring a source model trained on booster/preamp EDFAs (which incorporate internal EDFA features) to an ILA EDFA is non-trivial for the following reasons:
\begin{enumerate}
    \item Booster/preamp EDFAs are in-ROADM devices, where gain measurements are obtained via built-in OCMs. In contrast, ILAs typically lack integrated OCMs; their gain spectrum measurements are collected using external OCMs. Although these external variables are accounted for during data processing, they introduce additional noise.
    \item SS-NN models for booster/preamp EDFAs achieve optimal performance when trained with additional internal features. However, these features may not be available for ILAs due to vendor constraints. While it is possible to transfer a source model trained without these extra features, doing so would require maintaining dual models or sacrificing performance on the source EDFA.
\end{enumerate}

TL in these cases suffer from non-significant difference in source and target feature distributions, as well as a feature mismatch. This often leads to significant changes in model's input representations, requiring more measurements for domain adaptation. We utilize a covariance matching technique for TL - Correlation Alignment for Deep Domain Adaptation~(CORAL)~\cite{coral, coral2} in order to align the feature representations between source and target models. Essentially, CORAL loss aligns the second order statistics (i.e. covariance matrices) of the source and target domain for increased domain adaptability. The heterogeneous TL process is implemented as follows:

\begin{enumerate}
    \item \textbf{Reference Covariance Computation:}  
    From the training set of the source EDFA model, 128 randomly selected measurements are sampled. The output of the last hidden layer (with \(d=100\) neurons) is recorded; let this be represented by the matrix \(\mathbf{F}_S \in \mathbb{R}^{N \times d}\), where \(N=128\). The covariance matrix \(\mathbf{C}_S\) is then computed as:
    \begin{equation}
    \mathbf{C}_S = \frac{1}{N-1} \left( \mathbf{F}_S - \bar{\mathbf{F}}_S \right)^\top \left( \mathbf{F}_S - \bar{\mathbf{F}}_S \right),
    \end{equation}
    where \(\bar{\mathbf{F}}_S\) is the mean feature vector over the batch. This matrix is saved as a fixed reference.

    \item \textbf{Feature Imputation:}  
    For the target EDFA, any missing internal \ac{VOA} features are imputed with \(-999\) to ensure consistency in the model architecture.

    \item \textbf{Differential Fine-Tuning:}  
    The source EDFA model is fine-tuned using \(m\) additional measurements from the target EDFA. A differential learning rate is applied: the output layer is updated with a larger learning rate of \(1\times10^{-2}\), while each hidden layer is assigned a learning rate that is 10\% of that of the subsequent layer. Moreover, the learning rate for each layer is halved every 2000 epochs.

    \item \textbf{Loss Function Update:}  
    During fine-tuning, for each batch the target model's last hidden layer features \(\mathbf{F}_T \in \mathbb{R}^{N \times d}\) are obtained and their covariance matrix \(\mathbf{C}_T\) is computed similarly. From Eq.~\ref{loss_function}, the overall loss function~$\mathcal{L}_k$ for the \(k^\text{th}\) measurement is defined as:
    \begin{equation}
    \label{loss_function_coral}
    \mathcal{L}_k = \frac{1}{\sum_{i=1}^{95} c_i^k} \sum_{i=1}^{95} c_i^k \left[ g_{\text{pred}}^k(\lambda_i) - g_{\text{meas}}^k(\lambda_i) \right]^2 + \lambda \cdot \frac{1}{4d^2} \left\| \mathbf{C}_S - \mathbf{C}_T \right\|_F^2,
    \end{equation}
    where \(\|\cdot\|_F\) denotes the Frobenius norm and \(\lambda\) is the weighting factor for the CORAL loss.

    \item \textbf{Training:}  
    Finally, the target EDFA model is trained with the updated loss function for 10,000 epochs, using \(\lambda=0.4\). A base learning rate~($\alpha_l$) of $\alpha_l=1e-02$ was found to perform better in this case, as compared to $\alpha_l=1e-03$ used in homogeneous TL. 
\end{enumerate}

\subsubsection{Model Performance}

We first investigate the number of additional measurements required to effectively transfer a source \ac{EDFA} model to a target \ac{EDFA}. The additional measurements for fine-tuning are randomly loaded and fully-loaded channel configurations. To evaluate this, we compare the transfer learning (TL) performance of our proposed SS-NN model incorporating the CORAL loss with two alternative TL configurations:

\begin{enumerate} \item \textbf{TL with a Source Model Trained Without Additional Features:}
In this case, the source model is trained using only the basic features, excluding the internal \ac{EDFA} measurements. This configuration serves as a baseline for TL when the additional internal features are not available.
\item \textbf{TL with a Source Model Trained With Additional Features (MSE Loss):}  
Here, the source model is trained with the complete set of internal features, and TL is performed using the standard mean squared error (\ac{MSE}) loss.
\end{enumerate}

It should be noted that the direction of transfer introduces fundamentally different challenges. In the (Booster/Preamp)$\rightarrow$ILA TL scenario, there is an \emph{information loss} problem: the source booster/preamp models are trained with additional internal features, which are missing in the target \ac{ILA} dataset. In contrast, the ILA$\rightarrow$(Booster/Preamp) TL represents a case of \emph{information gain}—the source \ac{ILA} model is trained with imputed additional features, and is then transferred to a dataset with a complete feature set. 

Figure~\ref{ila_tl_mae} shows the mean absolute error (\ac{MAE}) of TL models as a function of the number of additional measurements per target gain setting used during training. The results, aggregated across all combinations of \acp{EDFA} in the COSMOS and Open Ireland testbeds under both random and goalpost channel configurations, indicate that heterogeneous TL requires more data points than homogeneous TL, which needs only a single fully loaded gain spectrum measurement. Moreover, the performance of a full-feature set SS-NN model degrades even with a large number of additional measurements, exhibiting negative TL. In contrast, models employing CORAL loss achieve performance comparable to those trained with a reduced feature set. Notably, the \ac{MAE} is higher for (Booster/Preamp)$\rightarrow$ILA TL than for ILA$\rightarrow$(Booster/Preamp) TL. We attribute this asymmetry to the greater difficulty of \emph{unlearning} the pre-trained source representations under a conservative training mechanism; optimizing this directional model asymmetry will be a focus of our future work.

Based on these results, we selected 32 additional measurements per target gain setting for ILA$\rightarrow$(Booster/Preamp) TL and 48 additional measurements for (Booster/Preamp)$\rightarrow$ILA TL. Figure~\ref{ila_tl_cdf} presents the cumulative distribution functions (CDFs) for both TL scenarios across all \acp{EDFA} in the COSMOS and Open Ireland testbeds under random and goalpost allocations. Models employing the modified loss function with CORAL loss perform similarly—and in some cases better—than the benchmark models. Specifically, for ILA$\rightarrow$(Booster/Preamp) TL, the CORAL loss yields superior performance with an absolute error of \(\leq\)0.22 dB across all test sets. For (Booster/Preamp)$\rightarrow$ILA TL, while models using CORAL loss outperform the benchmarks, overall performance is lower compared to other TL combinations, particularly under goalpost channel configurations.

These results demonstrate that incorporating CORAL loss into the SS-NN framework can enhance TL performance across \acp{EDFA} with different feature sets—eliminating the need to train and maintain multiple versions of the same model. Homogeneous TL is achieved with minimal additional measurements, while heterogeneous TL—particularly in the (Booster/Preamp)$\rightarrow$ILA scenario—requires a larger dataset to overcome information loss and feature mismatches. The proposed approach mitigates negative transfer effects but also yields consistent performance as evidenced by improved error distributions and lower MAE values. These findings provide a basis for further investigation into covariance alignment techniques for transfer learning in optical networks, especially in multi-span field networks where multiple vendor technologies are used.

\section{Conclusions and Future Work}
\label{section5}

In this paper, we introduced a generalized few-shot transfer learning architecture for EDFA gain spectrum modeling that integrates internal amplifier features within a Semi-Supervised Self-Normalizing Neural Network (SS-NN). The proposed two-phase training process, combining unsupervised pre-training with supervised fine-tuning, enables the SS-NN model to achieve high accuracy with limited labeled data. Furthermore, by incorporating a covariance matching (CORAL) loss, our approach effectively addresses domain discrepancies in heterogeneous transfer learning scenarios, particularly between booster/preamplifier EDFAs and ILAs. Experimental results on the COSMOS and Open Ireland testbeds show that the proposed framework not only reduces the required measurement overhead but also yields improved error distributions and lower mean absolute errors compared to state-of-the-art benchmarks. 

\added{Building on these findings, we aim to pursue several open challenges in the future. First, although an initial 10-month follow-up suggests that gain spectrum drift in the COSMOS \acp{EDFA} is below 0.1 dB~\cite{wangOpenEDFAGain2023}, a multi-year data-set is required to quantify the impact of \ac{EDFA} aging effects on model performance. Additionally, systematic exploration in the \ac{SHB} region, especially under extreme channel configurations could improve model reliability in power excursion scenarios. Finally, with the rise of flex-grid systems such as \ac{OSaaS}~\cite{osaas_ecoc_conf}, our future work will also focus on adapting these fixed-grid models to flex-grid operation.} 

\section*{Funding}
Supported by grants from Science Foundation Ireland (SFI): 12/RC/2276\_p2, 18/RI/5721, and 22/FFP-A/10598 and National Science Foundation (NSF): CNS-1827923, OAC-2029295, CNS-2112562, and CNS-2330333.

\bibliography{references}
  
\end{document}